\documentclass[journal]{IEEEtran}
\IEEEoverridecommandlockouts
\usepackage{cite}
\usepackage{makecell}
\usepackage[utf8x]{inputenc}
\usepackage{graphicx}
\usepackage{subfigure}
\usepackage{array}
\usepackage{booktabs}
\usepackage{url}
\usepackage[misc]{ifsym}
\usepackage[dvips]{color}
\usepackage{mathrsfs}
\usepackage[linesnumbered,ruled,vlined]{algorithm2e}
\usepackage{amsmath}
\usepackage{nccmath}
\usepackage{amsfonts,amssymb,bm}
\usepackage{amsthm}
\usepackage[dvipsnames]{xcolor}
\usepackage[noend]{algpseudocode}
\usepackage{overpic}
\usepackage{multirow}

\SetCommentSty{mycommfont}
\SetKwInput{KwInput}{Input}                
\SetKwInput{KwOutput}{Output}              

\usepackage[normalem]{ulem}
\useunder{\uline}{\ul}{}

\makeatletter
\renewcommand{\@pnumwidth}{1.75em}
\renewcommand{\@tocrmarg}{2.75em}
\makeatother

\newcolumntype{C}[1]{>{\centering\let\newline\\\arraybackslash\hspace{0pt}}m{#1}}

\begin{document}
\title{Towards Fine-Grained Indoor Localization based on Massive MIMO-OFDM System: Experiment and Analysis}
\author{Chenglong~Li,
		Sibren~De Bast,~\IEEEmembership{Student~Member,~IEEE,}
		Emmeric~Tanghe,~\IEEEmembership{Member,~IEEE,}\\
		Sofie~Pollin,~\IEEEmembership{Senior~Member,~IEEE}
		and Wout~Joseph,~\IEEEmembership{Senior~Member,~IEEE}
\thanks{This work is supported in part by the Excellence of Science (EOS) project MUlti-SErvice WIreless NETworks (MUSE-WINET), by the Research Foundation Flanders (FWO) SB Ph.D. fellowship under Grant 1SA1619N, and by the FWO project under Grant G098020N.}
\thanks{C. Li, E. Tanghe, and W. Joseph are with the WAVES group, Department of Information Technology, Ghent University-imec, 9052 Ghent, Belgium (e-mail: chenglong.li@ugent.be).}
\thanks{S. De Bast and S. Pollin are with the Department of Electrical Engineering, KU Leuven, 3000 Leuven, Belgium.}
}

\maketitle
\begin{abstract}
Fine-grained indoor localization has attracted attention recently because of the rapidly growing demand for indoor location-based services (ILBS). Specifically, massive (large-scale) multiple-input and multiple-output (MIMO) systems have received increasing attention due to high angular resolution. This paper presents an indoor localization testbed based on a massive MIMO orthogonal frequency-division multiplexing (OFDM) system, which supports physical-layer channel measurements. Instead of exploiting channel state information (CSI) directly for localization, we focus on positioning from the perspective of multipath components (MPCs), which are extracted from the CSI through the space-alternating generalized expectation-maximization (SAGE) algorithm. On top of the available MPCs, we propose a generalized fingerprinting system based on different single-metric and hybrid-metric schemes. We evaluate the impact of the varying antenna topologies, the size of the training set, the number of antennas, and the effective signal-to-noise ratio (SNR). The experimental results show that the proposed fingerprinting method can achieve centimeter-level positioning accuracy with a relatively small training set. Specifically, the distributed uniform linear array obtains the highest accuracy with about 1.63-2.5-cm mean absolute errors resulting from the high spatial resolution.
\end{abstract}

\begin{IEEEkeywords}
Massive multiple-input and multiple-output (MIMO), indoor localization, fingerprinting, multipath components, channel state information (CSI), orthogonal frequency-division multiplexing (OFDM), machine learning.
\end{IEEEkeywords}

\section{Introduction}
\label{sec:introduction}
\IEEEPARstart{I}{ndoor} location-based services (ILBS) have become an essential part of smart Internet-of-things (IoT) to support the extensive location-aware applications both in industry and social activities. Different from outdoor positioning, of which the precise location is usually provided by global navigation satellite systems (GNSS), numerous indoor localization solutions have been proposed based on different sensors and radio frequency (RF) platforms over the past decades. These include ultra-wideband (UWB), wireless fidelity (WiFi), radio frequency identification (RFID), Bluetooth, and multimodal-based systems, based on the requirement on the accuracy, coverage, and applicability \cite{GuSurvey2019}. Being available in most RF measurement devices, the received signal strength indicator (RSSI) characterizes the attenuation of radio propagation and has been widely utilized for indoor localization. Although RSSI (model-based or fingerprinting-based) can achieve meter-level positioning accuracy, it suffers from unreliable performance due to the multipath effect and dynamic scenarios. Assisted by multiple antennas or frequencies, several commercial off-the-shelf (COTS) IoT devices can provide a relatively accurate estimation of angle-of-arrival (AoA) or time-of-flight (ToF). For instance, the UWB-based system \cite{Dotlic2017} and WiFi-based systems \cite{SpotFi2015}. Exploiting both angle-based (AoA) and distance-based (RSSI or ToF) metrics, \cite{Hanssens2018,Li2019IPIN} have investigated the positioning accuracy of hybrid metrics and discussed the placement of RF devices based on the coverage requirement and positioning performance of the estimated metrics. Moreover, instead of utilizing geometric features (distance and angle), the received phase has also been used for accurate indoor localization due to its robustness to complex environments \cite{Zhu2015,Li2019,Li2020}.
\par
Unlike RSSI is the energy accumulation at the medium access control (MAC) layer, channel response as a physical (PHY) layer metric can characterize how RF signal propagates from the transmitter to the receiver. Precisely measuring the wireless channel generally involves dedicated setups, such as vector network analyzer (VNA) or software-defined radio (SDR), which challenge practical applications. But the advancement of RF hardware circuits results in an increasing number of COTS devices, such as Intel WiFi link 5300 NIC \cite{Halperin2010,FILA2012}, enabling channel response collection. Exploiting channel state information (CSI), some pioneer works have achieved sub-meter indoor positioning accuracy, even in cluttered scenarios \cite{FILA2012,Xiao2012,Souvik2012}. In \cite{SpotFi2015}, a prototype SpotFi was established, which estimated the AoA via a two-dimensional (2-D) multiple signal classification (MUSIC) algorithm and localized the user based on the AoA from the direct path. In \cite{ChenCSI2020}, similar to SpotFi, AoA and ToF were estimated from the calibrated CSI based on an improved MUSIC algorithm. Together with CSI amplitude-based fingerprints, an AoA-enhanced probabilistic fingerprinting method was established. Instead of estimating AoAs from CSI, \cite{Wang2016,Song2018} proposed to utilize the CSI (amplitude or phase) directly to implement the fingerprinting system, which also achieved sub-meter median positioning accuracy.
\par
Massive (or large-scale) multiple-input multiple-output (MIMO) systems do not only benefit communications in terms of channel capacity and spectral efficiency, but also have the potential for accurate localization due to the high angular resolution. In \cite{Hu2014}, a closed-form estimating signal parameters via rotational invariance technique (ESPRIT) was proposed for the incoherently distributed sources. To address the positioning problem of the narrow-band massive MIMO system, especially in a multipath-rich scenario, \cite{Garcia2015} proposed to jointly process snapshots of several distributed arrays based on compressed sensing without line-of-sight (LoS) distinguishing and achieve sub-meter accuracy with high probability. \cite{Savic2015,Decurninge2018} proposed the fingerprint-based localization for cellular massive MIMO systems based on machine learning. In \cite{Arnold2019,CNNMaMIMO2020}, massive MIMO channel sounder prototypes were built for indoor use cases. The CSI data was fed to deep convolutional neural networks (CNN) to train the positioning model and achieved sub-meter \cite{Arnold2019} and centimeter-level \cite{CNNMaMIMO2020} accuracy, respectively. Moreover, instead of training from scratch, transfer learning was introduced in \cite{CNNMaMIMO2020} to reduce the modeling time, which relieved CNN's dependence on the scenario variety to some extent, and made CNN-based solution more practical.
\par
In this paper, we focus on the fine-grained indoor localization based on a massive MIMO system with a standard cellular bandwidth and propose to localize the user through the channel components. The main contributions of this paper are as follows.
\begin{enumerate}
\item We have established a massive MIMO indoor localization testbed supporting PHY-layer metric collection, namely, channel state information. Different antenna topologies, including uniform linear array (ULA), distributed ULA (DIS), and uniform rectangular array (URA), have been investigated.
\item We have calibrated the CSI offsets across frequency and antenna domain due to synchronization and hardware errors. On top of that, the multipath components (MPCs) have been extracted and analyzed based on the space-alternating generalized expectation-maximization (SAGE) algorithm.
\item We have implemented a fingerprinting system based on the extracted MPCs of direct links. The corresponding positioning performance has been evaluated in cases of different metrics, antenna topologies, sizes of the training set, number of antennas, and the effective signal-to-noise ratio (SNR).
\item The proposed MPCs-based fingerprinting method is generalized from the perspective of propagation. It requires less training set than the available deep learning-based massive MIMO indoor positioning solutions (e.g., CNN-based method\cite{CNNMaMIMO2020}).
\end{enumerate}
\par
The remainder of this paper is organized as follows. Section-II introduces the system setting of the massive MIMO campaign and details of the CSI collection. In Section III, the CSI calibration in the frequency and antenna domain is investigated. In Section IV, the MPCs are extracted and analyzed based on the calibrated CSI. Exploiting the MPCs of direct links, the regression-based fingerprinting system is established. In Section V, the positioning performance is evaluated and analyzed. Furthermore, potential future directions are discussed. Section VI concludes this paper.
\par
\textit{Notation}: Vectors and matrices are denoted by lower case boldface letter $\bf{a}$ and upper case boldface letter $\bf{A}$, respectively. The set of $M\times N$ real and complex matrices are denoted by $\mathbb{R}^{M\times N}$ and $\mathbb{C}^{M\times N}$, respectively. $j=\sqrt{-1}$ is the imaginary unit. $(\cdot)^{\top}$ represents the transpose operator, $(\cdot)^{\rm{H}}$ the conjugate transpose operator. $\odot$ denotes the Hardmard product operator. $\vert\cdot\vert$ and $\Vert\cdot\Vert$ denote the absolute value and $l_2$-norm operator, respectively. ${\rm{mode}}(\cdot)$ is the most frequent value finding operator. $\mathcal{F}^{-1}_N(\cdot)$ represent the $N$-point discrete inverse Fourier transform. Further, $Q(\cdot)$ and $Q^{-1}(\cdot)$ denotes the Q function and inverse Q function, respectively.

\begin{figure}[t]
\centering
\includegraphics[width=0.485\textwidth]{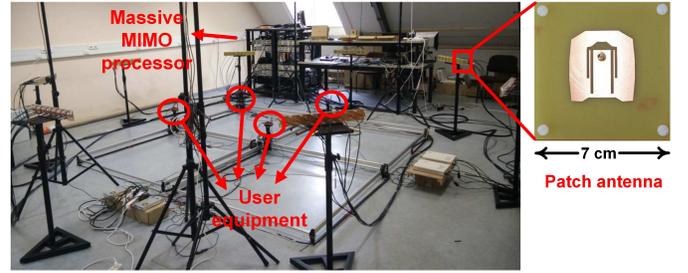}
\caption{Measurement campaign and setups of massive MIMO-OFDM prototype for the channel state information collection.}
\label{fig:SetupPics}
\end{figure}

\begin{figure}[t]
\centering
\includegraphics[width=0.485\textwidth]{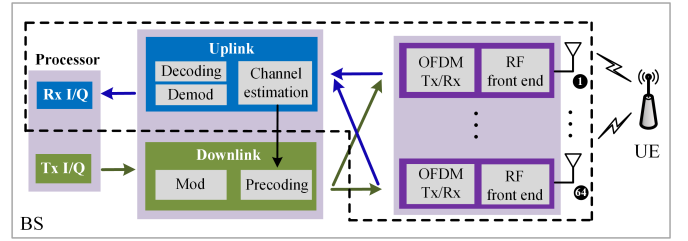}
\caption{Signal processing chain of the massive MIMO-OFDM testbed. The diagrams within the dashed region represent the uplink procedure adopted for positioning in the experiment.}
\label{fig:TestbedStruc}
\end{figure}
\section{Massive MIMO Experiment: Channel State Information Collection}
Based on the massive MIMO testbed at KU Leuven ESAT-TELEMIC \cite{CNNMaMIMO2020,guevara2020weave}, the single-cell channel response is collected under different kinds of antenna topologies. The involved massive MIMO system consists of a base station (BS), equipped with 64 patch antennas, and the four universal software radio peripherals (USRPs) with a single dipole antenna each, acted as the user equipment (UE). The measurement campaign of CSI collection is shown in Fig. \ref{fig:SetupPics}, which was conducted in the MIMO-lab of KU Leuven. The four dipole-antenna USRPs were deployed within a targeted area 3 m$\times$3 m. For this massive MIMO system, a time division duplex (TDD)-based frame structure has been adopted, and orthogonal frequency-division multiplexing (OFDM) modulation and demodulation have been performed via the Xilinx field-programmable gate array (FPGA). During the measurement, the BS was under the control of LabVIEW Communications MIMO Application Framework \cite{NIreport2019}, which allows performing CSI measurement between the BS and the UE. The BS (all 64 antennas) receives the orthogonal pilots sent by the four UEs simultaneously and conducts the channel estimation. The signal processing chain is presented in Fig. \ref{fig:TestbedStruc}. For the CSI collection, only the uplink procedure has been considered as the dashed region in Fig. \ref{fig:TestbedStruc} shown. The pilot tone consists of 100 sub-carriers, which are evenly spaced in frequency. Therefore, the measured CSI of a single transmission can be represented by the complex matrix,
\begin{equation}\label{eq:CSI_H}
\mathbf{H}_{\rm{CSI}} = \left\lbrace H_{n_r,n_k}\right\rbrace\in\mathbb{C}^{64\times100},
\end{equation}where $n_r\in\lbrace1,2,\cdots,64\rbrace$, $n_k\in\lbrace1,2,\cdots,100\rbrace$ represent the index of antenna and sub-carrier, respectively. The center frequency of the massive MIMO prototype is 2.61 GHz and bandwidth 20 MHz. The transmitted power is 15 dBm. 
\par
CSI-based WiFi for decimeter-level positioning \cite{FILA2012,Xiao2012,Souvik2012,SpotFi2015,ChenCSI2020} inspires us to increase the accuracy further (cm- or mm-level accuracy) by deploying the large antenna array. One promising use case of indoor massive MIMO localization is the intelligent surface \cite{Hu2017}. In this case, the large antenna array can be embedded into the walls, which extends massive MIMO for fine-grained positioning and sensing, making the physical environment interactive. Moreover, another possible application is the accurate positioning of automated robots in factories or warehouses. The massive MIMO-based system can support multiple-robot localization simultaneously, which is also little affected by the poor visibility compared with the vision-based solutions. Different from the UWB-based indoor localization \cite{GuSurvey2019,Kulmer2017}, which also aims for cm-level accuracy, the established massive MIMO system has adopted a standard cellular signal bandwidth and the OFDM scheme, which is compatible with the current long-term evolution (LTE), sub-6 GHz 5G, and WiFi communications systems.

\begin{figure}[t]
\centering
\setlength{\abovecaptionskip}{-0.01cm}
\includegraphics[width=0.485\textwidth]{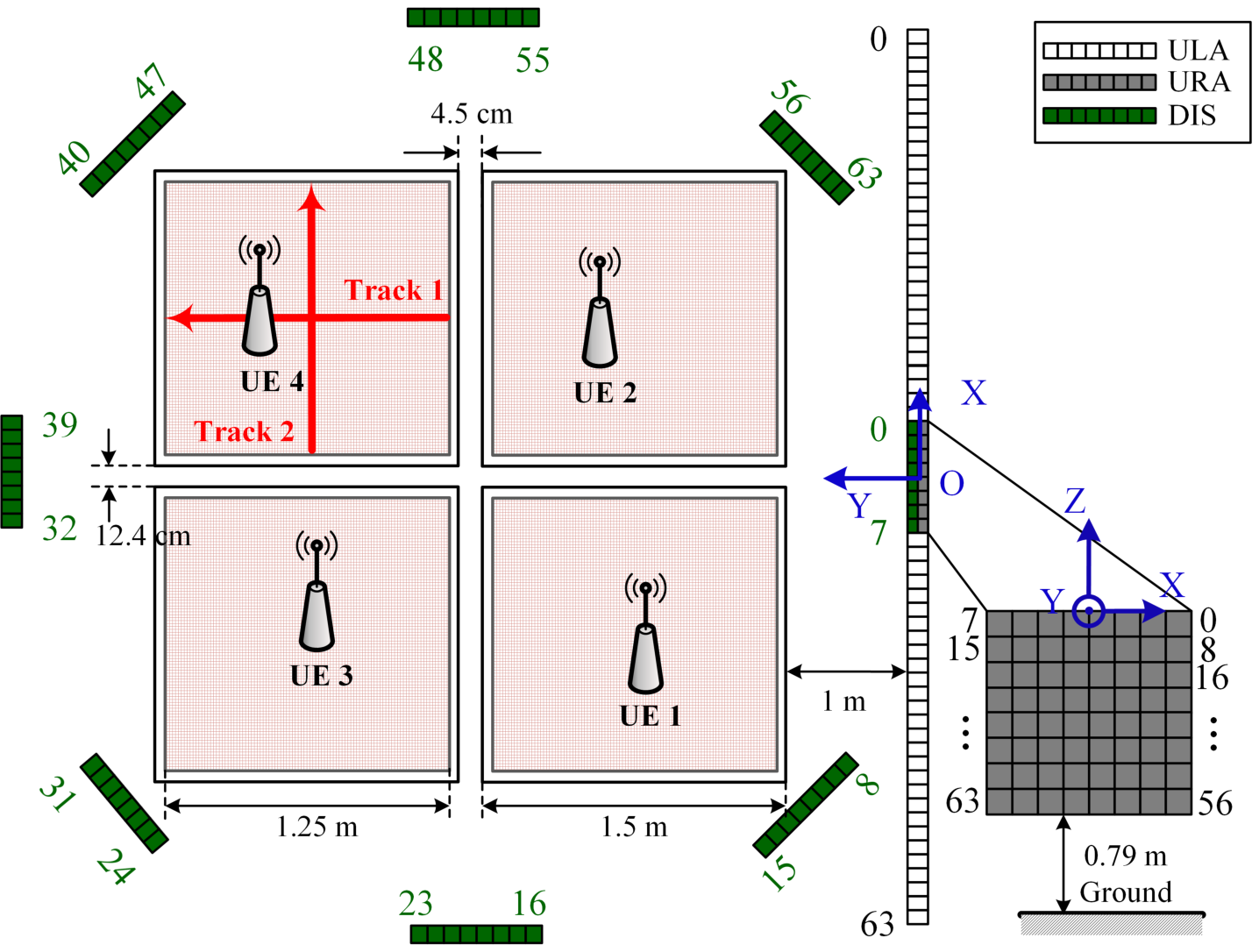}
\caption{Massive MIMO measurement campaign of CSI collection. Different antenna topologies are included: uniform linear array (ULA), distributed ULA (DIS), and uniform rectangular array (URA).}
\label{fig:MAMIMOSetup}
\end{figure}
\par
The massive MIMO testbed has been designed for the flexible deployment of the antenna arrays. It provides three topologies of the antenna array for CSI collection, specifically, a uniform linear array (ULA) of $1\times64$ antennas, a uniform rectangular array (URA) of $8\times8$ antennas, and eight distributed ULAs of $1\times8$ antennas (DIS). The spacing between the adjacent antenna elements is 7 cm. The height of both ULA and DIS is one meter, while the lowest antenna elements of URA are located 79 cm above the floor. The height of UE is 40 cm. The ULA and URA are deployed on the one side of the targeted area, where the DIS surrounds the UEs, as shown in Fig. \ref{fig:MAMIMOSetup}. The numbers alongside the antenna elements are the order of CSI in antenna domain (e.g., $n_r$ in (\ref{eq:CSI_H})). The UEs are moved by the computerized numerical control (CNC) X-Y table along a zigzagged trajectory, which guarantees the collected ground truth of UE with less than 1-mm error. The moving stride of UE is 5 mm, so we have collected 252004 CSI samples in total for the four UEs. The dataset is publicly available \cite{CSIdatasets} to encourage further research.  

\section{Channel State Information Calibration}
\begin{figure}[t]
\centering
\subfigure[UE within the calibration set]{
\includegraphics[width=0.238\textwidth]{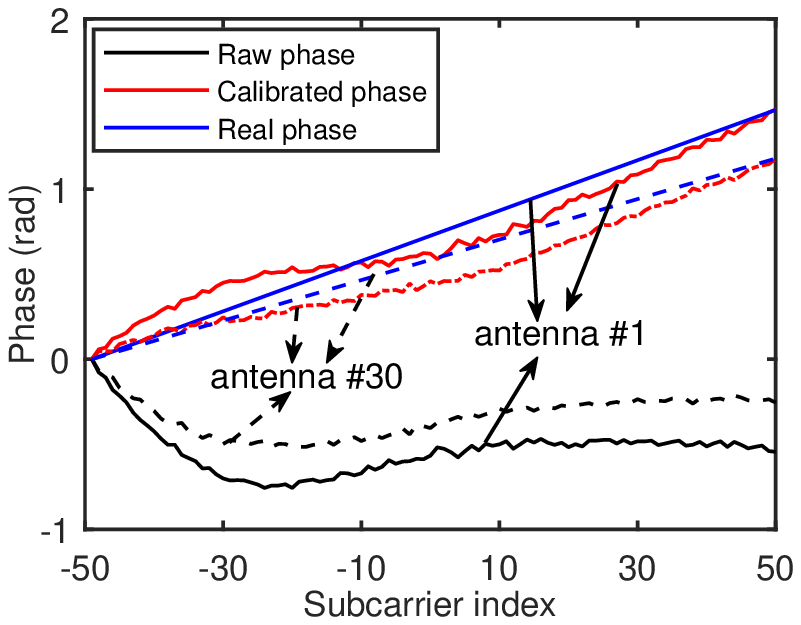}
}
\hspace{-4.9mm}
\subfigure[UE outside the calibration set]{
\includegraphics[width=0.238\textwidth]{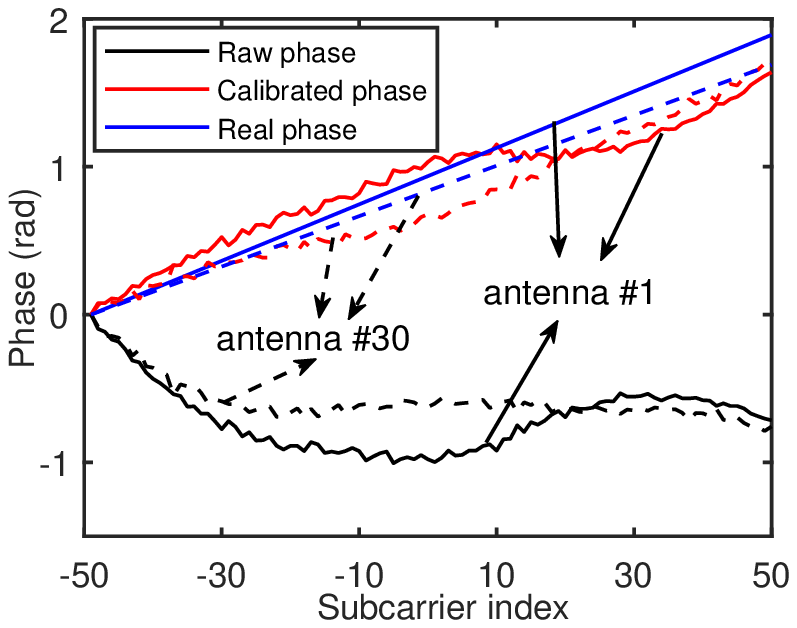}
}
\hspace{-4.9mm}
\subfigure[UE within the calibration set]{
\includegraphics[width=0.238\textwidth]{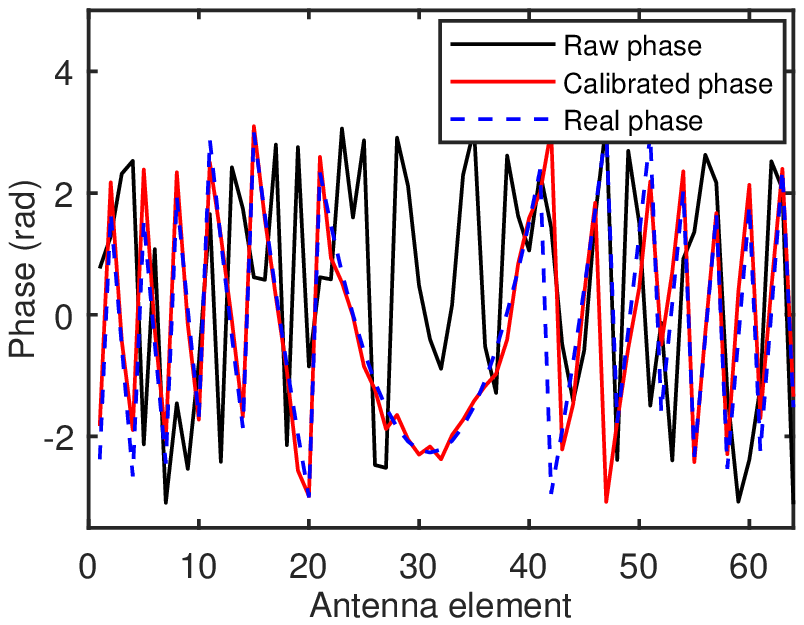}
}
\hspace{-4.9mm}
\subfigure[UE outside the calibration set]{
\includegraphics[width=0.238\textwidth]{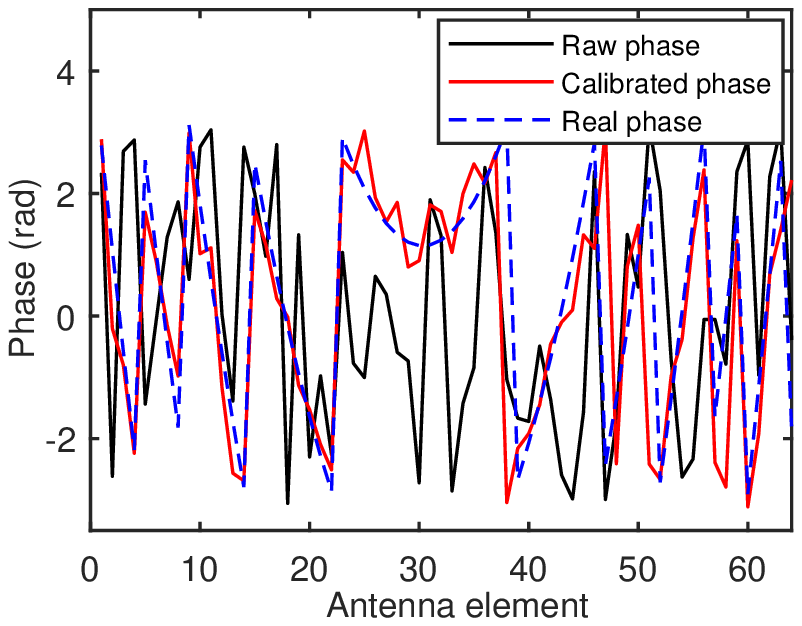}
}
\caption{The phases of the raw CSI and calibrated CSI compared with the real phase calculated via the ground truth: (a)-(b) along the sub-carrier (aligned by the first subcarrier), (c)-(d) along the antenna array.}
\label{fig:PhaComp}
\end{figure}

As a fine-grained PHY layer metric, CSI depicts the rich channel characteristics via the acute phase reacting in time, frequency (multi-frequency), and space (multi-antenna) domain. However, when conducting channel estimation under the assumption of perfect time and frequency synchronization between the transceiver, the actual phase in the measured CSI will be contaminated by the synchronization errors. To this end, the received phases suffer from the sampling frequency offset (SFO), and symbol timing offsets (STO), etc. \cite{SpethOFDM1999,TadayonCSI2019}. According to \cite{Tarighat2005,ZhuSplicer2018}, the implementation of orthogonal frequency division multiplexing (OFDM)-based PHY layer is susceptible to the effect of in-phase and quadrature-phase (IQ) imbalance in the front-end analog processing, which may cause nonlinear phase distortion on CSI. Due to the random initial phase generated by the local voltage-controlled oscillator and imperfect compensation of the phase-locked loop, carrier phase offsets (CPO) are imposed on the received phases. In Figs. \ref{fig:PhaComp}(a)-(b), the phases extracted from the raw CSI of two antenna elements (\#1 and \#30 of ULA) are shown as the black (dash) lines. In theory, the phase should increase monotonically with the increasing frequency as the blue (dash) lines of the real phases are presented. But due to the phase distortion mentioned above, the phases of the measured CSI present erroneous tendency. Moreover, for the UE in front of the antenna array (e.g., ULA), the phase received by each antenna element should decrease first and then increase, as well as the distance, and reach the minimum when the distance between the UE and antenna element is shortest. After being wrapped between $-\pi$ and $\pi$, the real phase can be represented by the blue dash lines in Figs. \ref{fig:PhaComp}(c)-(d), whereas the phases of the raw CSI are arbitrary. It is because we did not calibrate across the antenna elements before the CSI collection. Even though we have adopted the same type of antenna and equally long cables, there are possible phase offsets across antennas due to hardware heterogeneity.
\par

\subsection{Calibration in Frequency Domain}
Overall, the measured CSI in the presence of the aforementioned phase errors can be given by,
\begin{equation}\label{eq:CSI}
\begin{aligned}
\!\tilde{H}_{n_r,n_k}\!=\!&\left(\sum_{l=1}^L H_{n_r,n_k}^{(l)}\right)\!\cdot\!\exp\!\left(\!-\!j\!\left(\zeta_{\rm{SFO}}^{(n_k)}\!+\!\zeta_{\rm{STO}}^{(n_k)}\!+\!\zeta_{\rm{IQ}}^{(n_k)}\right)\right)\!\!\\
&\quad\quad\quad\quad\cdot\exp\left(\!-j\left(\eta_{\rm{CPO}}+\xi_{\rm{ant}}^{(n_r)}\right)\right)\!+\omega,
\end{aligned}
\end{equation}where the first item on the right side is the desired multipath channel response. $L$ is the number of propagation paths, and $\omega$ the complex noise in the CSI matrix. $\zeta_{\rm{SFO}}$, $\zeta_{\rm{STO}}$, and $\zeta_{\rm{IQ}}$ represent the phase shift caused by SFO, STO, and IQ imbalance, respectively. The SFO phase shift is proportional to the subcarrier index, which can be mitigated through a multiple linear regression across subcarriers \cite{TadayonCSI2019,SpotFi2015}. Due to the cyclic-shifting of channel impulse response (CIR), STO results in the high amplitude peaks at the far end of the power delay profile (PDP) \cite{TadayonCSI2019}, which can be utilized to calibrate the STO. So the most frequent far-end peak index among multiple packets can be obtained by,
\begin{equation}\label{eq:STO}
\hat{K}_{\rm{STO}}=\mathop {\rm{mode}}\left(\mathop {\arg\max}\limits_{{n_k}}\!\vert\mathcal{F}^{-1}_{100}(H_{n_r,n_k})\vert^2\right),n_k>1,
\end{equation}where $n_k>1$ means that the first peak of the PDP generally is the LoS link. So the estimated STO can be expressed as $\zeta_{\rm{STO}}^{(n_k)}=\frac{n_k\hat{K}_{\rm{STO}}}{100}$ \cite{TadayonCSI2019}, where 100 is the number of subcarriers. But we can observe that for a specific $\hat{K}_{\rm{STO}}$, $\zeta_{\rm{STO}}$ also presents a linear relationship with the subcarrier index. To this end, we also can utilize linear regression to estimate the STO. 
\par
\begin{figure}[t]
\centering
\setlength{\abovecaptionskip}{-0.02cm}
\includegraphics[width=0.485\textwidth]{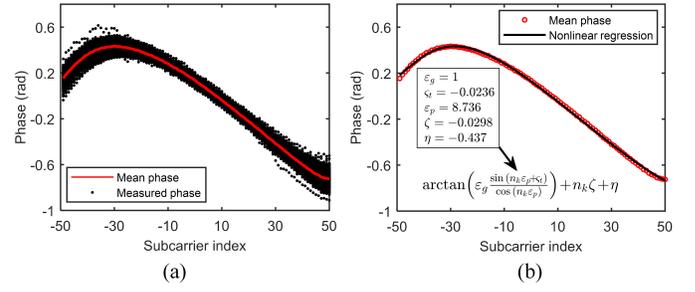}
\caption{(a) Residual phase offsets of 500 CSI samples after removing the LoS channel response and the corresponding mean residual phase. (b) Nonlinear fitting results on the mean residual phase.}
\label{fig:NLfitting}
\end{figure}

According to the experimental results in \cite{ZhuSplicer2018}, the nonlinear phase shift caused by IQ imbalance is quite stable along the time scale but sensitive to the frequency. Fig. \ref{fig:NLfitting}(a) presents 500 samples of the residual measured phase after removing the LoS channel response based on the ground truth. It can be observed that the phase shifts caused by IQ imbalance for different samples present a similar tendency along the subcarrier index, which means the sampling location has little impact on the calculation of IQ imbalance. To this end, we can calibrate the IQ imbalance by a nonlinear fitting on 
\begin{equation}\label{eq:IQ}
\zeta_{\rm{IQ}}^{(n_k)}=\arctan{\left(\varepsilon_g\frac{\sin{(n_k\varsigma_t+\varepsilon_p)}}{\cos{(n_k\varsigma_t)}}\right)},
\end{equation}where $\varepsilon_g$, $\varepsilon_p$ represent the gain and phase mismatch, respectively. $\varsigma_t$ denotes the unknown time offset. 
\par
The phase offset caused by CPO is represented by $\eta_{\rm{CPO}}$ in \eqref{eq:CSI}, which can be regarded as a random constant after the initiation of the transceiver \cite{ChenCSI2020}. In summary, we can calibrate the phase shifts caused by SFO, STO, CPO, and IQ imbalance through the following nonlinear regression,
\begin{equation}\label{eq:NLregression}
\mathop {\arg\min}\limits_{{\bf{\Upsilon}}}\!\sum\limits_{n_k}\left(\Delta\bar{\Theta}_{n_k}\!-\!\zeta_{\rm{IQ}}^{(n_k)}\!-\!n_k\zeta_{\rm{SFO/STO}}\!-\!\eta_{\rm{CPO}}\right)^2,
\end{equation}where ${\bf{\Upsilon}}=[\varepsilon_g,\varsigma_t,\varepsilon_p,\zeta_{\rm{SFO/STO}},\eta_{\rm{CPO}}]$. $\Delta\bar{\Theta}$ denotes the average residual phase offset after removing the LoS channel response based on the ground truth of the transceiver. The regression problem \eqref{eq:NLregression} can be solved via the Levenberg–Marquardt algorithm. Fig. \ref{fig:NLfitting}(b) shows the nonlinear fitting results and the involved parameters for the CSI calibration. After the above frequency domain calibration, the relation between the phase and frequency has been recovered. The red (dash) lines in Figs. \ref{fig:PhaComp}(a)-(b) present the calibrated phases. Besides the UE within the calibrated set, we also used the obtained calibration parameter set ${\bf{\Upsilon}}$ to calibrate the measurements outside the calibrated set, as shown in Fig. \ref{fig:PhaComp}(b), which validates the effectiveness of the proposed phase calibration method.

\subsection{Calibration in Antenna Domain}
Moreover, as mentioned above, even though we utilized the same type of antenna and the same length of cables, there still exists phase offsets due to the heterogeneity of hardware components. As presented in \eqref{eq:CSI}, $\xi_{\rm{ant}}$ is constant along the frequency and time scale, but variant across antenna elements. Since the BS (all the receiver antennas) shares the same oscillator and local reference time, it is reasonable to assume that all the phase shifts, except for $\xi_{\rm{ant}}^{(n_r)}$, $n_r\in(1,2,\cdots,64)$, are constant along the antenna domain. It should be noted that the reported phases from the CSI are wrapped between $-\pi$ to $\pi$. To avoid the phase ambiguity towards the estimation of $\xi_{\rm{ant}}^{(n_r)}\in(-\pi,\pi]$, we estimate the antenna phase shift in the complex field, given by,
\begin{equation}\label{eq:ant}
\mathop {\arg\min}\limits_{{\xi_{\rm{ant}}^{(n_r)}}}\!\left\vert\sum\limits_{n_{\rm{RP}}}\!\sum\limits_{n_k}\!\left(\exp\big(j\Delta{\Psi}_{n_k,n_{\rm{RP}}}\big)\!-\!\exp\big(j\xi_{\rm{ant}}^{(n_r)}\big)\right)\right\vert^2,
\end{equation}where $\Delta{\Psi}$ is the residual phase after the nonlinear calibration mentioned above. $n_{\rm{RPs}}$ is the index of the reference points (RPs) for calibration. After the calibration in antenna domain, as presented in Fig. \ref{fig:PhaComp}(c), the calibrated phases match well with the real phase. Similarly, we also utilized the UE outside the calibrated set to validate the calibration effect. As shown in Fig. \ref{fig:PhaComp}(d), the phases along the antenna elements have been calibrated and aligned despite some small fluctuations.
\subsection{Note to Practitioners}
To calibrate the phase in the measured CSI, we select a small part of the CSI dataset merely acting as the calibration set to mitigate the nonlinear phase shifts and antenna phase offsets. After this, we use the obtained ${\bf{\Upsilon}}$  and $\xi_{\rm{ant}}$ to calibrate the newly measured CSI directly without the above extensive calibration procedures. Notably, the number of CSI samples in the calibrated set should not be less than 64, which guarantees the problem in \eqref{eq:ant} is non-negative definite. For the fingerprinting system (in Section IV), the reference points for the calibration are self-contained (i.e., the grids of fingerprints).
\begin{figure}[t]
\centering
\subfigure[AoAs: using raw CSI]{
\includegraphics[width=0.238\textwidth]{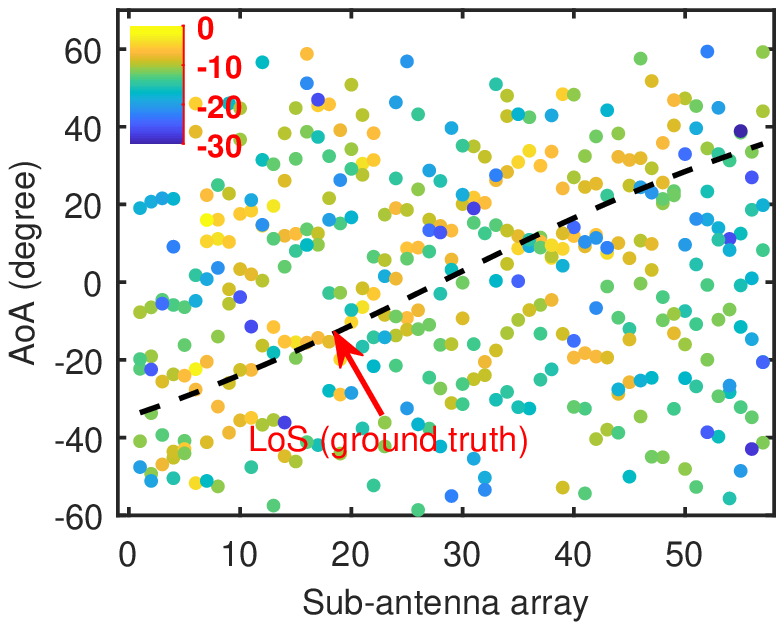}
}
\hspace{-4.9mm}
\subfigure[ToFs: using raw CSI]{
\includegraphics[width=0.238\textwidth]{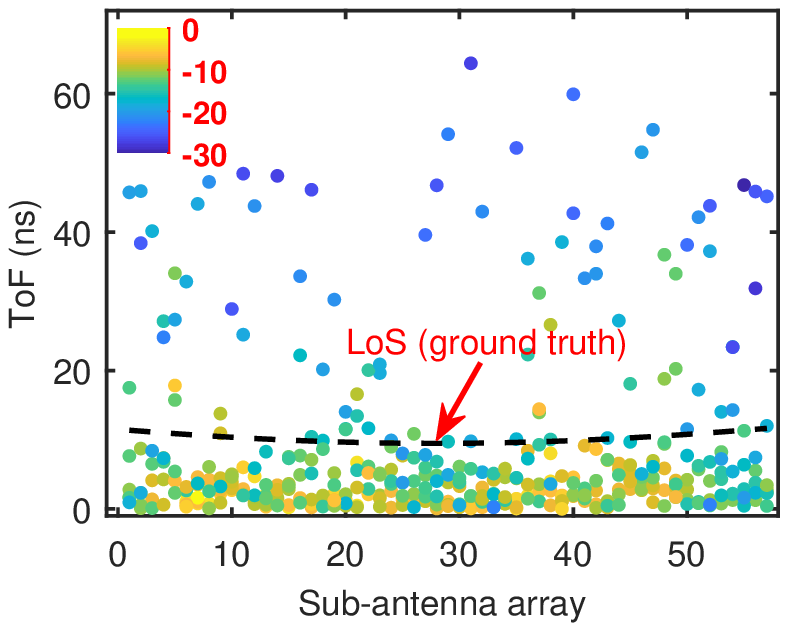}
}

\subfigure[AoAs: after CSI calibration]{
\includegraphics[width=0.238\textwidth]{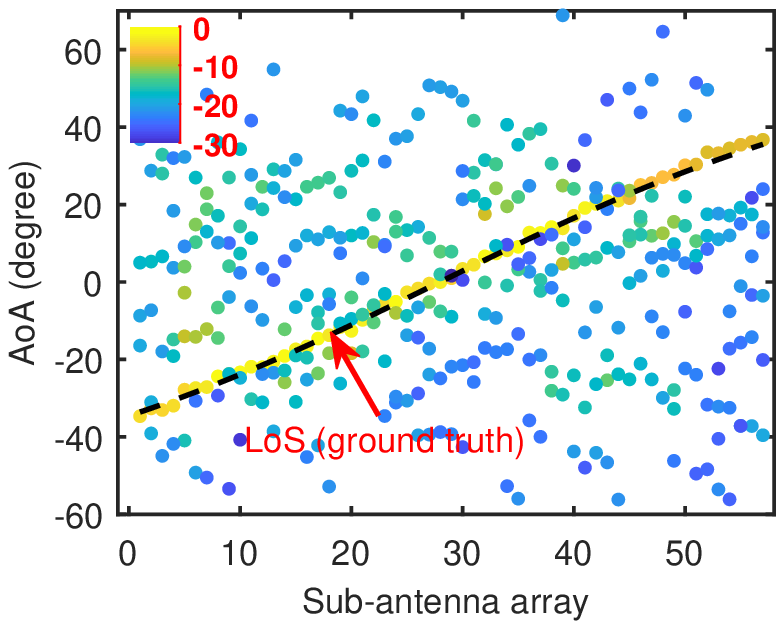}
}
\hspace{-4.9mm}
\subfigure[ToFs: after CSI calibration]{
\includegraphics[width=0.238\textwidth]{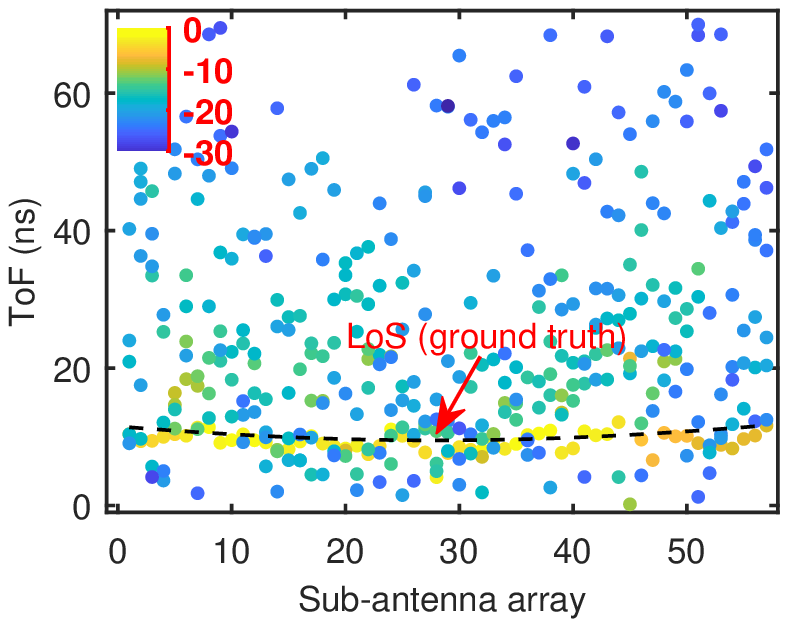}
}
\caption{Multipath components of ULA using the 8-element sliding sub-antenna arrays. The colorbar denotes the normalized power of each multipath component in dB.}
\label{fig:MPCresults}
\end{figure}
\section{Fingerprinting System Design}
Fingerprinting-based positioning techniques localize the UE by comparing its location-related metric(s) to a predefined radio map on the targeted area. For fingerprint matching, there are generally two mapping solutions: classification and regression. For the regression scheme, only the trained matching model is stored in the radio map. This saves lots of resources compared to the classification scheme in which all reference samples need to be stored in the radio map. Therefore, a regression scheme has been adopted in this paper. Unlike the CSI depicting the channel between the transceiver implicitly, the multipath components (extracted by, e.g., SAGE algorithm) characterize the propagation from the perspective of power and geometry (angle and distance), making the channel model more explainable. This section will focus on the analysis of the MPCs, the discussion of antenna array partition, and the design of the fingerprinting system based on the MPCs of direct links (namely, amplitude, AoA, and ToF).

\begin{figure*}[t]
\centering
\setlength{\abovecaptionskip}{-0.02cm}
\subfigure[]{
\includegraphics[width=0.328\textwidth]{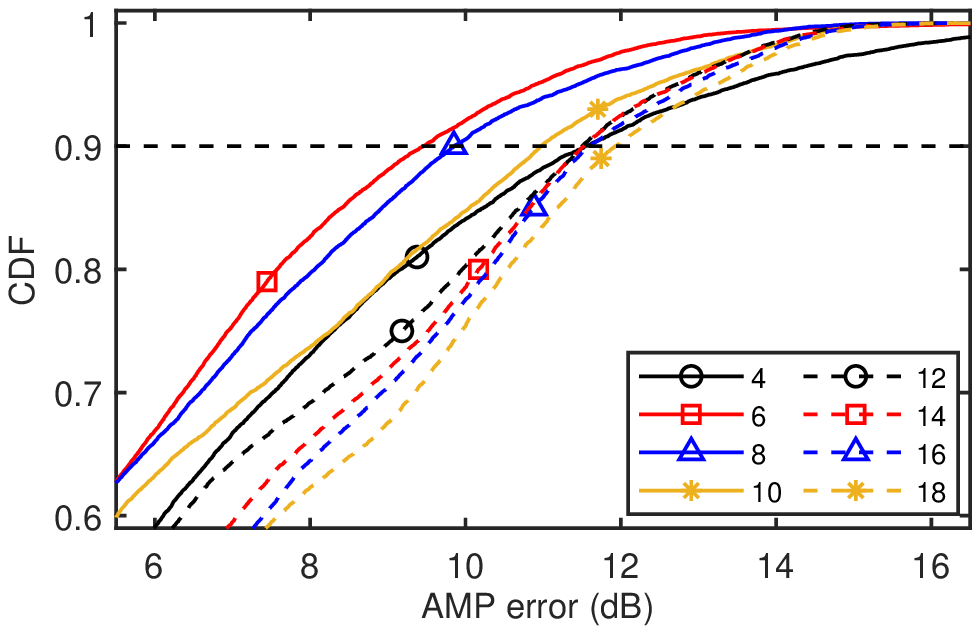}
}
\hspace{-4.5mm}
\subfigure[]{
\includegraphics[width=0.328\textwidth]{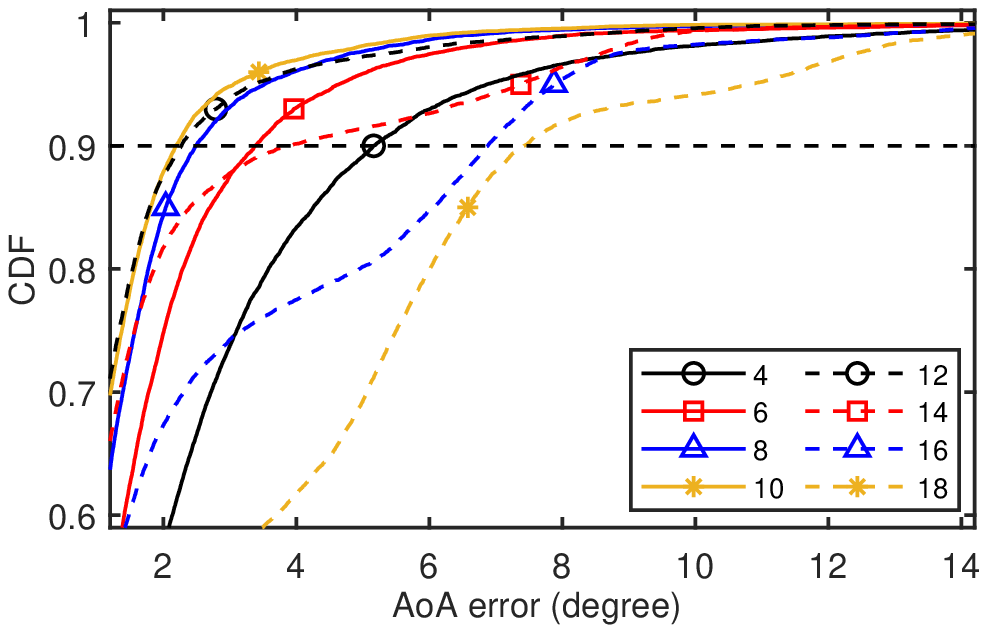}
}
\hspace{-4.5mm}
\subfigure[]{
\includegraphics[width=0.328\textwidth]{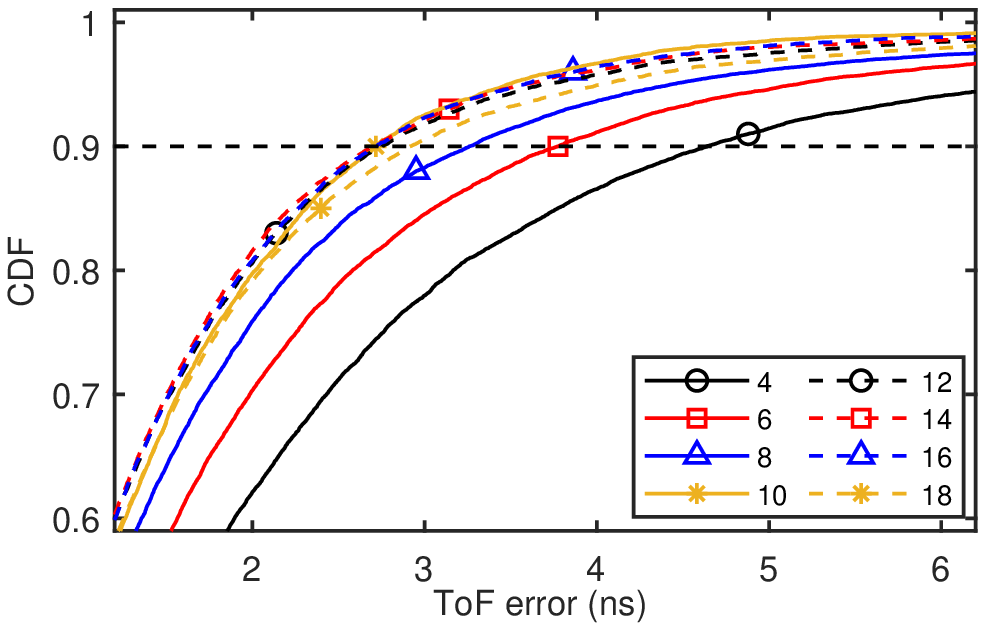}
}
\caption{Impact of number of antennas in sub-array: CDF of (a) AMP, (b) AoA, and (c) ToF for estimation errors. AMP represents the amplitude of MPCs.}
\label{fig:subAntSize}
\end{figure*}

\begin{figure*}[t]
\centering
\setlength{\abovecaptionskip}{-0.02cm}
\subfigure[]{
\includegraphics[width=0.328\textwidth]{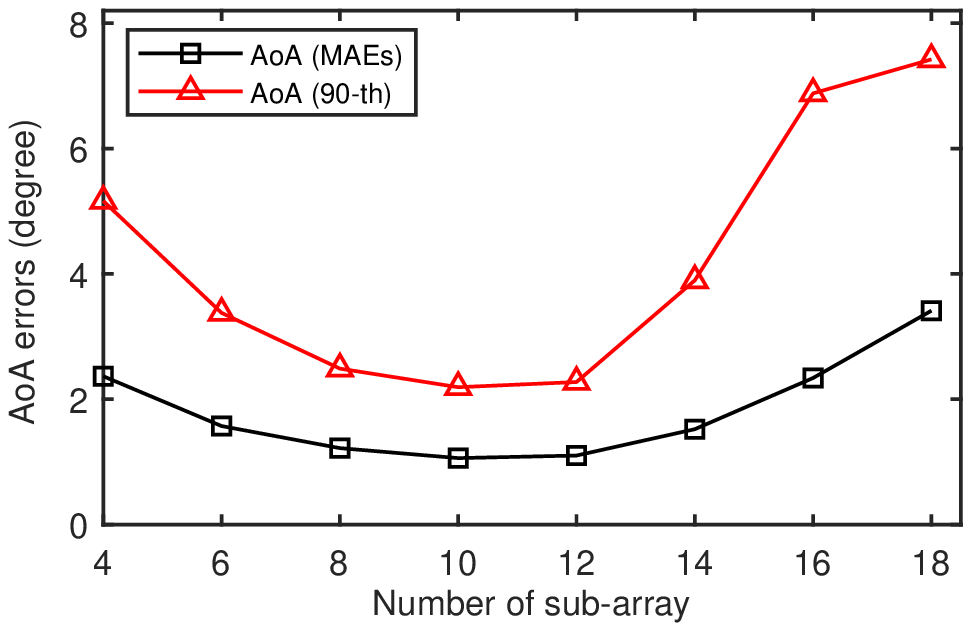}
}
\hspace{-4.5mm}
\subfigure[]{
\includegraphics[width=0.328\textwidth]{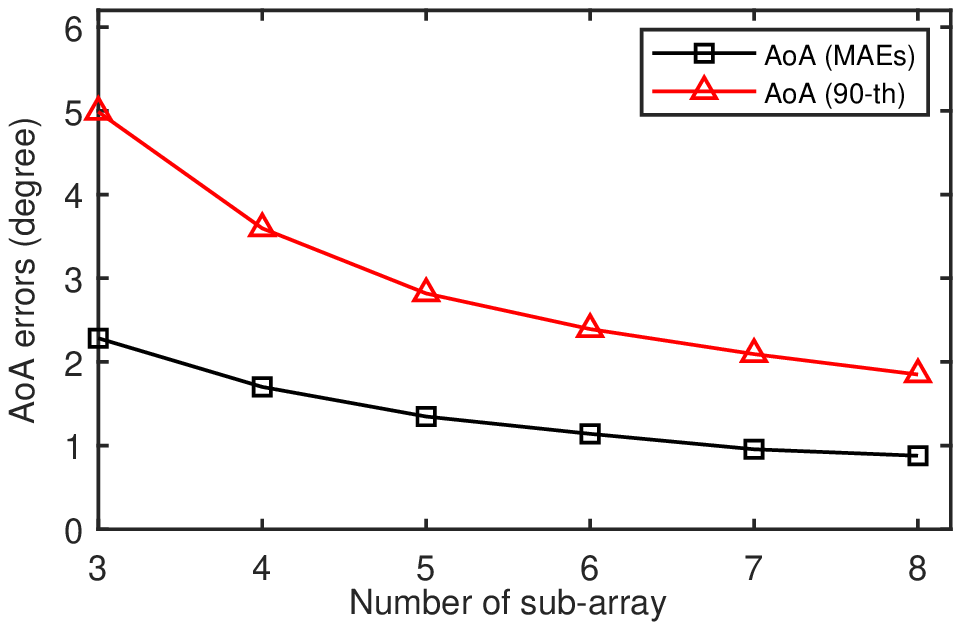}
}
\hspace{-4.5mm}
\subfigure[]{
\includegraphics[width=0.328\textwidth]{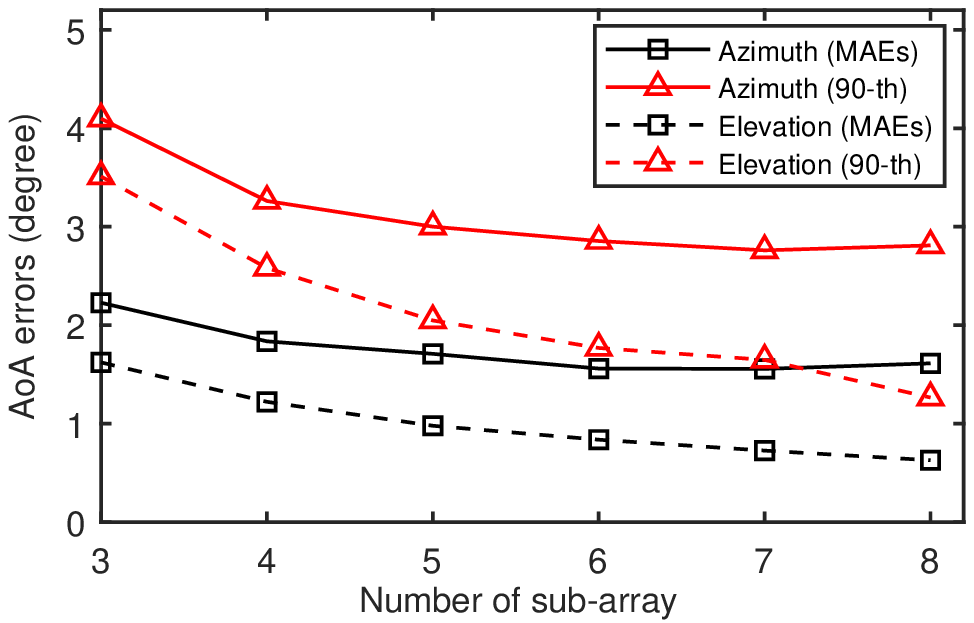}
}
\caption{Impact of number of antennas on AoA estimation accuracy: (a) ULA, (b) DIS, (c) URA.}
\label{fig:AoA_subarray}
\end{figure*}

\subsection{Multipath Components Extraction}
Under assumption of a far-field and time-invariant MIMO channel, the channel transfer function at the $k$-th frequency bin $f_k$, ($1\leq k\leq K$) is given by
\begin{equation}\label{eq:channel}
{\bf{H}}(f_k)\!=\!\sum_{l=1}^L\!\alpha_l c_{\rm{Rx}}({\bf{\Omega}}_{{\rm{Rx}},l}) c_{\rm{Tx}}^{\top}({\bf{\Omega}}_{{\rm{Tx}},l})\exp(\!-j2\pi f_k\tau_l),\!
\end{equation}where $\alpha_l, \tau_l$ denote the amplitude and time of flight (ToF) of the $l$-th multipath, respectively. $c_{i}({\bf{\Omega}}_{i,l}), (i={\rm{Tx},\rm{Rx}})$ is the steering vector at transceiver. The directional vector ${\bf{\Omega}}_{i,l}$ is uniquely determined by the spherical coordinates $(\theta_{i,l},\varphi_{i,l})\in[-\frac{\pi}{2},\frac{\pi}{2}]\times[0,\pi]$, namely, ${\bf{\Omega}}_{i,l}=[\sin\theta_{i,l}\cos\varphi_{i,l},\sin\theta_{i,l}\sin\varphi_{i,l},\cos\theta_{i,l}]^{\top}$, where $\theta_{i,l},\varphi_{i,l}$ represent the elevation angle of arrival (EAoA), azimuth angle of arrival (AAoA), respectively. In this paper, we only consider the angle at the receiver side because the UE's antenna is a single dipole. To extract the specular MPCs ${\bf{\Theta}}=[\alpha_l,\theta_l,\varphi_l,\tau_l]$ from the CSI, we introduce a computationally efficient frequency domain (FD) SAGE algorithm, which updates channel parameters sequentially with low-dimensional maximization steps \cite{FDSAGE2002}. The iterations stop when the power of a single path is 30 dB below the peak path power. We can determine the number of paths $L$ in this way. Besides E-step and M-step above, the initialization of ${\bf{\Theta}}$ is also related with algorithm performance and convergence. According to \cite{SAGE1999,FDSAGE2002}, ToF and AoA can be initialized through frequency/spatial correlation, etc. Moreover, for the ULA, there is no information about elevation. So for the ULA and DIS topologies in our experiment, only the azimuth is available. In this case, the $c_{\rm{Rx}}({\bf{\Omega}}_{{\rm{Rx}},l})$ in \eqref{eq:channel} can be simplified as $c_{\rm{Rx}}(\varphi_l)\!=\left\lbrace\exp{(\!-j\frac{2\pi d}{\lambda}(m-1)\sin\varphi_l)}\right\rbrace\in\mathbb{C}^{M\times1}$, $m=1,2,\cdots,M$, where $d$ is the spacing between the adjacent antenna, and $\lambda$ the wavelength. The M-step for ULA should be modified correspondingly. We refer to \cite{FDSAGE2002} for the detailed algorithm.

\subsection{Antenna Array Partition}
As the prerequisite of \eqref{eq:channel} presented, the FD-SAGE algorithm is under the far-field assumption, which generally requires the distance between the transceiver, or the distance from scatterer to the transceiver is smaller than the Rayleigh distance, defined by $2D^2/\lambda$, where $D$ is the largest dimension of the antenna array. Therefore, the Rayleigh distance for the ULA (1$\times$64), DIS (1$\times$8), and URA (8$\times$8) in our experiment are 349.35 m, 5.46 m, and 9.25 m, respectively, which may be not easy to satisfy for indoor scenarios. To this end, the large antenna array is divided into multiple successive sub-arrays by a sliding window to apply the FD-SAGE algorithm. Fig. \ref{fig:MPCresults} presents the MPCs results of the ULA without or with CSI calibration using the 8-element sliding sub-array. It can be observed that there are clear LoS-trajectories of AoA and ToF along the sub-antenna arrays after the calibration. The LoS components are distinguished clearly by the higher path power. Moreover, in Fig. \ref{fig:MPCresults}(c), the AoAs of the LoS path present distinct angle offsets along the antenna array, which verifies the spherical wavefront of the large antenna array.
\par
To determine the sub-array size, we specify the number of antennas by evaluating the estimated accuracy of the LoS channel components. Fig. \ref{fig:subAntSize} presents the cumulative distribution function (CDF) of absolute errors of amplitude, AoA, and ToF of the LoS link for the long ULA topology. The real AoA and ToF are calculated based on the ground truth of the transceiver, while the amplitude (in dB) is compared with the standard path loss. As shown in Fig. \ref{fig:subAntSize}, the amplitude requires a relatively small sub-array size to obtain the high estimation accuracy, whereas the ToF is the opposite. When converting the errors of amplitude, AoA, and ToF into spatial errors, it is reasonable to take the AoA accuracy as the benchmark to determine the size of the sub-array. For example, when the distance of the transceiver is 3 m, the 1-dB amplitude error, 1-degree AoA error, and 1-ns ToF error represent about $(10^{1/20}-1)\times3=$36.61 cm, $\pi/180\times3=$5.24 cm, and 29.98 cm, respectively. Fig. \ref{fig:AoA_subarray} presents the AoA estimation errors of different numbers of antennas in the sub-array in the case of three antenna topologies. For the ULA, as shown in Fig. \ref{fig:AoA_subarray}(a), when the number of antennas is between 8 and 12, the estimating accuracy is relatively stable (about 1.06$\sim$1.22-degree MAEs, and 2.19$\sim$2.49-degree 90th percentile errors). So we can set the size of the sub-array as 8$\sim$12 for ULA. Note that 8 is selected for the positioning performance evaluation in Section V. In Fig. \ref{fig:AoA_subarray}(b), the AoA errors decrease as increasing the number of antenna elements in the sub-array for the DIS topology. So we have not partitioned the distributed ULA further and kept the antenna array as 1$\times$8. For URA in Fig. \ref{fig:AoA_subarray}(c), the azimuth and elevation reach steady with slight fluctuations when the size of the sub-array ranges from 6 to 8. Therefore, we set the sub-array size as 6$\times$6, namely the multiple sub-arrays as 6$\times$6$\times$9, to guarantee more partitions and increase the spatial resolution of the URA.

\subsection{Fingerprinting System Design}
In the framework of fingerprinting regression, the objective is to find a nonlinear mapping between the input metrics $\mathcal{X}$ and the UE's coordinates $\mathcal{Y}$, namely $\mathcal{M}:\mathcal{X}\in\mathbb{R}^\mathfrak{D}\rightarrow\mathcal{Y}\in\mathbb{R}^2$ (2-D positioning), where $\mathfrak{D}$ is the number of the input features. The metric can be the amplitude, AoA, and ToF of the MPCs. In this paper, we introduce $\epsilon$-support vector regression ($\epsilon$-SVR) \cite{Smola2004} to conduct the nonlinear mapping, which allows at most $\epsilon$ deviation from the actual targets. To manage nonlinear regression, the kernels, such as the Gaussian kernel, have been utilized to map the input metric to higher dimensional feature space and then apply the standard $\epsilon$-SVR algorithm. It should be noted that only one output target is available for the regression in $\epsilon$-SVR, so we implement two SVR models for the 2-D positioning ($x$ and $y$ separately). The parameters involved in SVR, including the $\epsilon$-insensitive band, kernel scale, and regularization term strength, are automatically optimized through Bayesian optimization.

\begin{figure}[t]
\centering
\setlength{\abovecaptionskip}{-0.01cm}
\subfigure[Track 1]{
\includegraphics[width=0.237\textwidth]{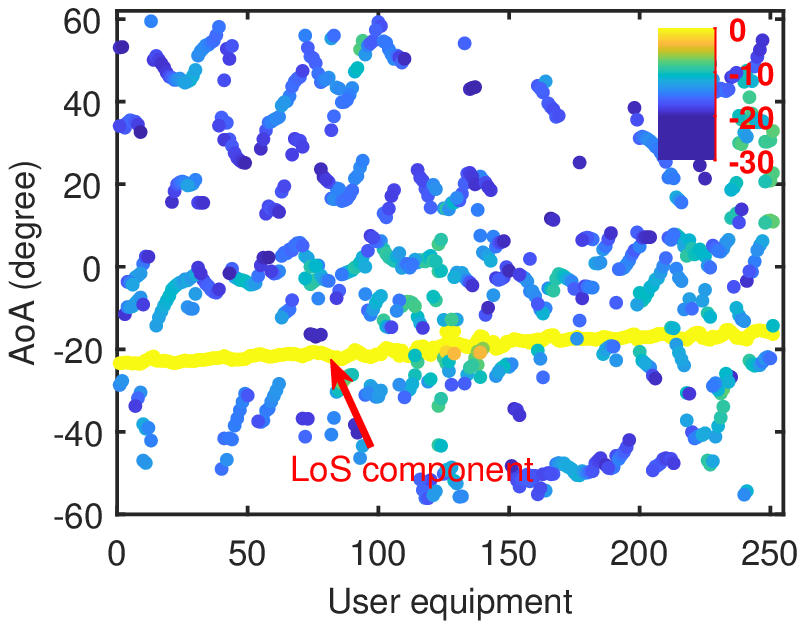}
}
\hspace{-4.9mm}
\subfigure[Track 1]{
\includegraphics[width=0.237\textwidth]{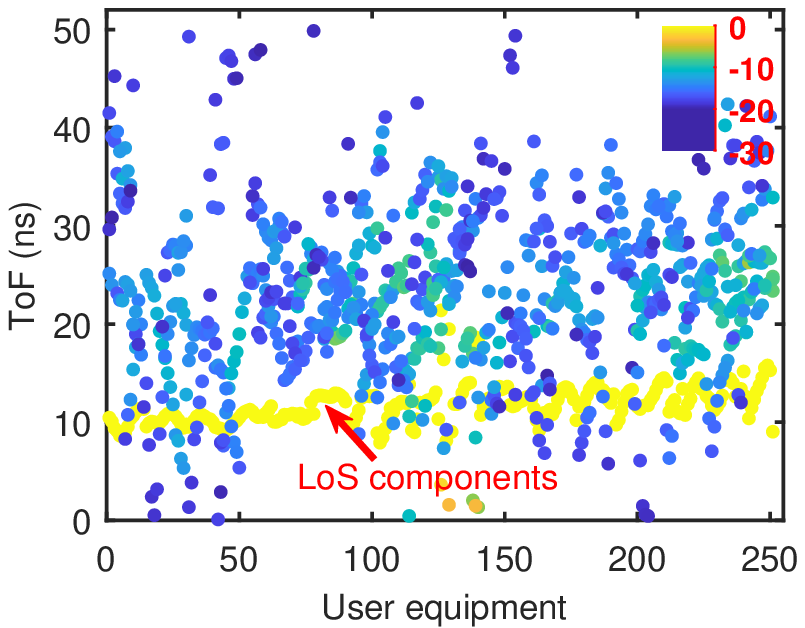}
}

\subfigure[Track 2]{
\includegraphics[width=0.237\textwidth]{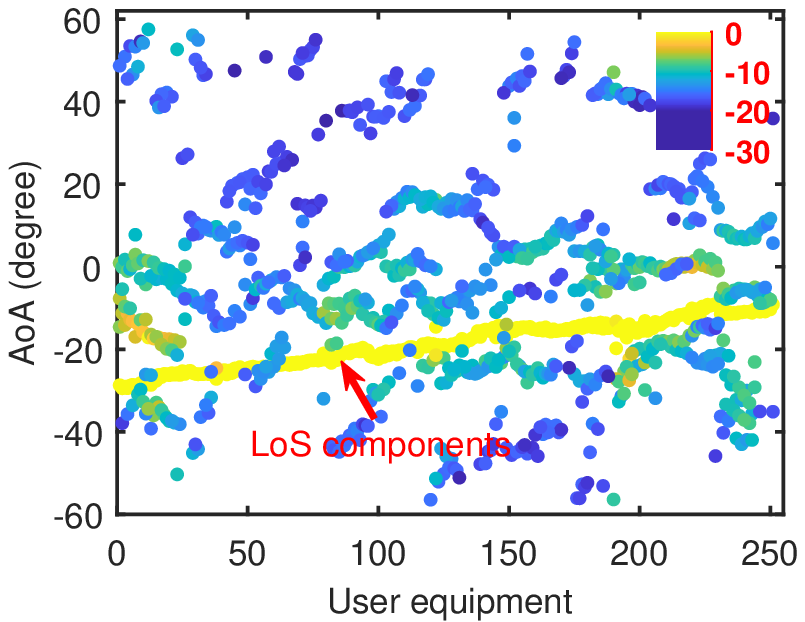}
}
\hspace{-4.9mm}
\subfigure[Track 2]{
\includegraphics[width=0.237\textwidth]{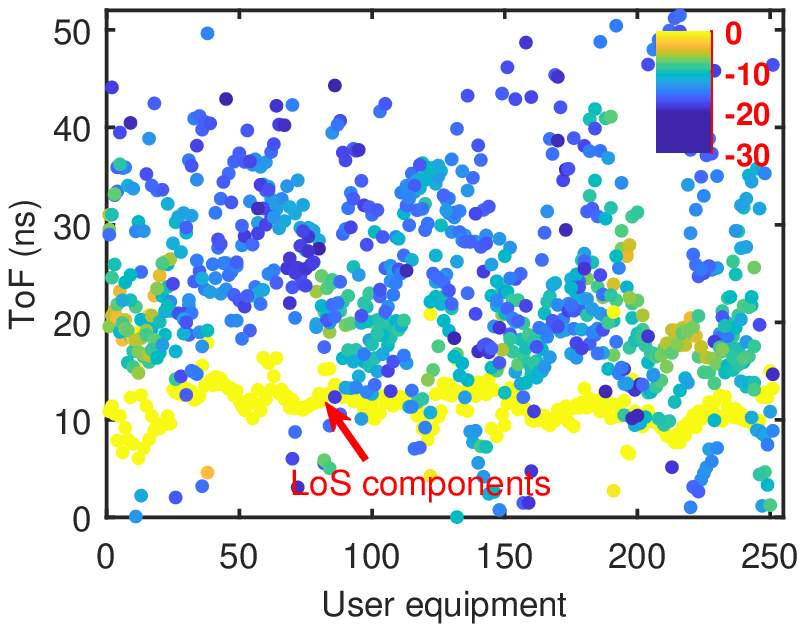}
}
\caption{Multipath components of the first subarray of ULA for track 1 and track 2 in Fig. \ref{fig:MAMIMOSetup} for multipath analysis.}
\label{fig:MPCresultsMU}
\end{figure}
\subsubsection{Discussion of Input Features}
Multipath-assisted localization has attracted attention by exploiting the geometrical metrics (distance or angle) of LoS and the multipath. Generally, given the prior knowledge of the floor plan \cite{Witrisal2016,Kulmer2017} or within the simultaneous localization and mapping (SLAM) framework \cite{Gentner2016,Ulmschneider2018,Leitinger2019}, the fixed but unknown virtual anchors' locations (the mirrored positions of the BS w.r.t. flat surfaces) can be obtained. To this end, the likelihood of the possible position can be enhanced with the assistance of the virtual anchors and the corresponding distance/angle-based components. Suppose there exist consistent MPCs associating with the unknown virtual anchor. In that case, the regression-based fingerprinting method has the intrinsic merit without the need of the floor plan or SLAM, since it does not require the positions of the (virtual) anchors but establishes the nonlinear mapping between the features and UEs' positions directly.
\par
Fig. \ref{fig:MPCresultsMU} present the MPCs results of two consecutive trajectories in Fig. \ref{fig:MAMIMOSetup} (namely, Track 1 and 2). Except for the LoS components, the other MPCs have no consistent or long-lifetime trajectory, which means it is not possible to associate a group of consistent MPCs with a fixed virtual anchor. The birth-to-death spans of the MPCs are very short (less than about 25 cm). It is because a consistent-MPCs trajectory generally requires the specular reflection from a long planar surface. If the surrounding is cluttered, most of the MPCs are from scattering or small flat surfaces. To this end, for a given channel order $L$, the virtual anchors' positions vary when the UE moves, making it impossible to establish a specific mapping between the features and locations. Note that, as presented in Fig. \ref{fig:SetupPics}, the measurement campaign was conducted in a LoS-dominated scenario despite the cluttered deployment outside the targeted area. So in this paper, only the MPCs from the direct links have been used for the fingerprinting design.
\begin{figure}[t]
\centering
\includegraphics[width=0.485\textwidth]{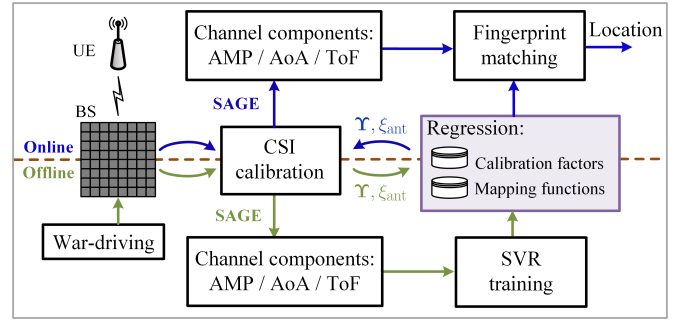}
\caption{The overall fingerprinting system architecture.}
\label{fig:SysArchitecture}
\end{figure}

\subsubsection{Fingerprinting Structure}
Fig. \ref{fig:SysArchitecture} shows the architecture of the proposed massive MIMO fingerprinting method. During the offline phase, the CSIs at the RPs are collected and calibrated using \eqref{eq:NLregression} and \eqref{eq:ant} in Section III. The involved parameters for CSI calibration are stored at the local database (radio map) for later online processing. After the calibration, the CSIs are fed to the maximum likelihood estimator (e.g., FD-SAGE) to extract the MPCs. As mentioned above, we only consider the components from the direct links. So the channel order can be set as one, which significantly reduces the computation time of FD-SAGE. Then the regression model, namely, $\epsilon$-SVR, has been adopted to establish the nonlinear mapping between the training metric (amplitude, AoA, or ToF) and the ground truth of RPs. The trained $\epsilon$-SVR models (for $x$- and $y$-coordinates prediction) are also stored at the local database. In the proposed fingerprinting system, besides the single metric, we also consider the hybrid scheme of these three metrics and the two-metric combinations. These hybrid metrics can be fed directly to the nonlinear SVR model, which will be assigned different weights at a higher dimensional feature space.
\par
During the online phase, the stored calibration factors are utilized directly for the newly measured CSI without the extensive optimization procedures of the offline phase. After that, the MPCs of the direct links are extracted using the FD-SAGE algorithm and fed to the trained SVR mapping models to obtain the 2-D location of the target.

\begin{figure*}[t]
\centering
\setlength{\abovecaptionskip}{-0.02cm}
\subfigure[ULA]{
\includegraphics[width=0.328\textwidth]{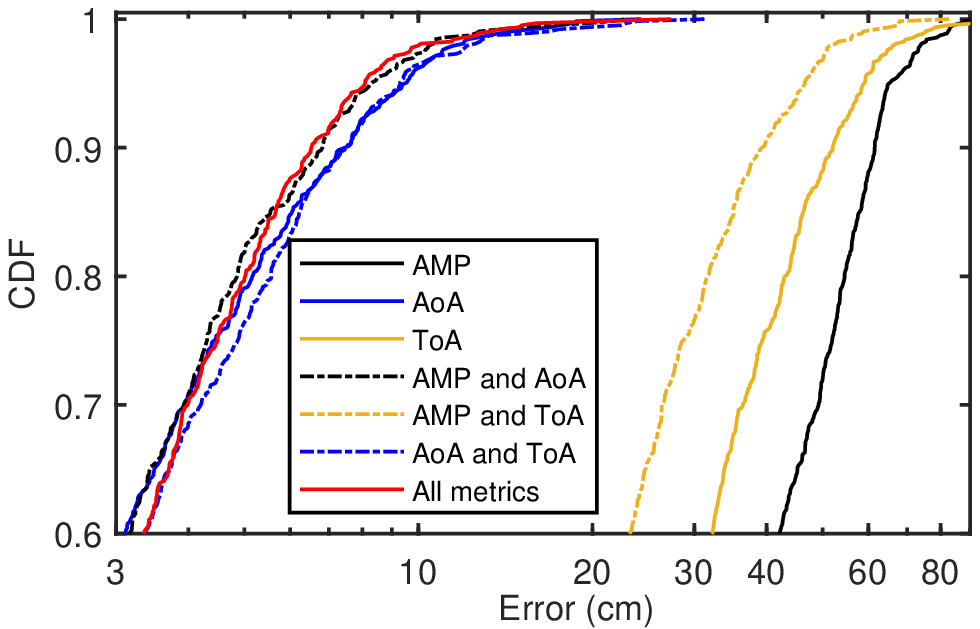}
}
\hspace{-4.5mm}
\subfigure[DIS]{
\includegraphics[width=0.328\textwidth]{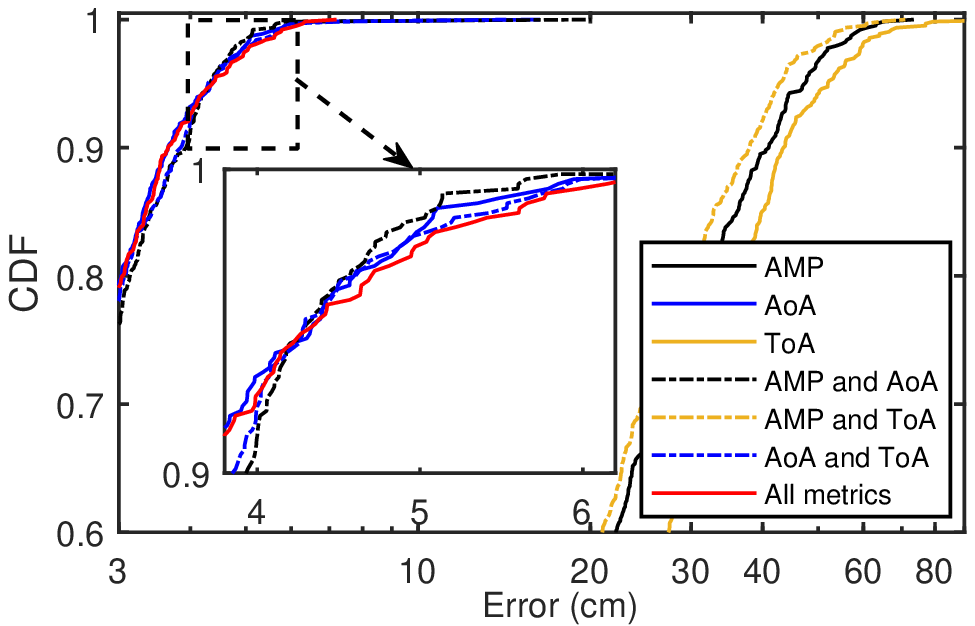}
}
\hspace{-4.5mm}
\subfigure[URA]{
\includegraphics[width=0.328\textwidth]{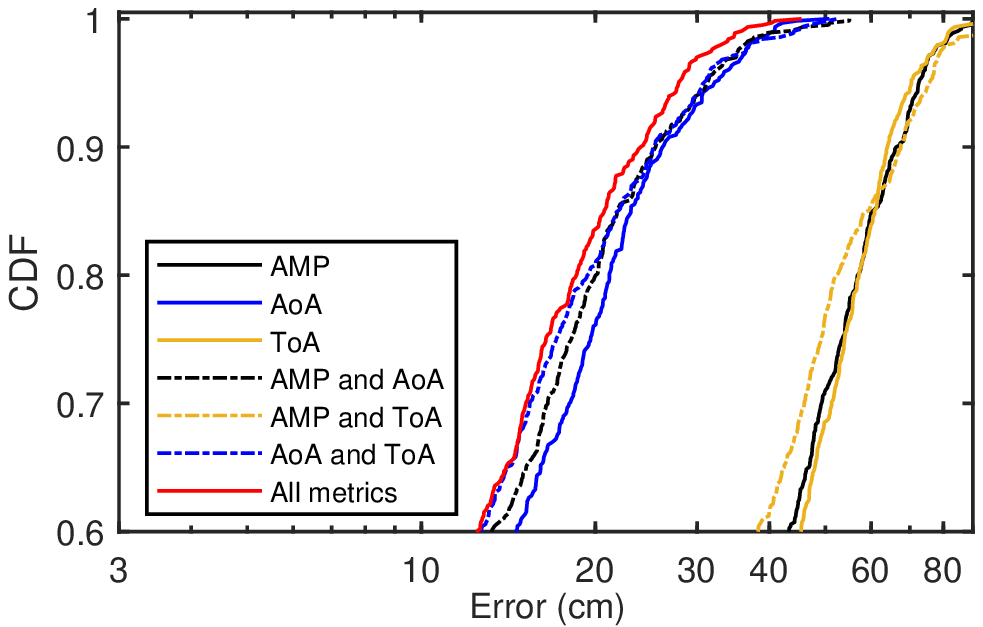}
}
\caption{Positioning accuracy of different single-metric and hybrid-metric schemes based on 11$\times$11-grid fingerprints.}
\label{fig:SVR_metrics}
\end{figure*}
\section{Performance Evaluation}
In this section, we investigate the positioning performance of the indoor massive MIMO localization system based on the collected CSI dataset. Instead of exploiting CSI directly, we have established the fingerprinting system based on the MPCs of the direct links (Section IV), which localizes the UEs from the perspective of propagation and geometry. We have adopted $N\times N$-grid fingerprints for each UE to train the model and the residual measurement to evaluate the positioning performance. In this section, the positioning accuracy of a single metric (amplitude, AoA, or ToF) and the hybrid metrics using the $\epsilon$-SVR regression model are evaluated. Moreover, the impacts of antenna topology, the number of antennas, and the effective SNR are presented and analyzed.
\subsection{Impact of Different Metrics}
Fig. \ref{fig:SVR_metrics} presents the CDF of the positioning errors in case of $N=11$, namely the resolution of grid is 12.5 cm. Regarding the three single-metric schemes (amplitude, AoA, and ToF), the AoA-based fingerprinting method distinctly outperforms the other two metrics, achieving about one-tenth of the positioning errors of ToF- (or amplitude-) based metric. Therefore, when involving AoA in the hybrid-metric schemes, AoA dominates the overall positioning accuracy. Comparing the AoA and all-metric based performance, as shown in Fig. \ref{fig:SVR_metrics}, ToF and amplitude only have slight improvement on ULA and URA topologies (about 0.5-2 cm). Compared to AoA-based accuracy, the all-metric case has no improvement but a little overfitting with 1-2 mm worsening for the DIS topology. However, when it comes to the hybrid amplitude-ToF scheme, the accuracy improves distinctly (about 2-10-cm median errors decreasing) than the single metric (amplitude or ToF). 

\begin{table}[t]
\centering
\caption{Percentile errors of $\epsilon$-SVR in the case of 11$\times$11-grid fingerprints in cm}
\begin{tabular}{c|c|c|c}
\hline
\multirow{2}{*}{\textbf{Errors}} & \multirow{2}{*}{\textbf{ULA}} & \multirow{2}{*}{\textbf{DIS}} & \multirow{2}{*}{\textbf{URA}} \\& & &             \\ \hline\hline
\textbf{Median}  & 2.67  & \textbf{1.88}  & 9.93  \\ \hline
\textbf{90-th}   & 6.51  & \textbf{3.65}  & 23.94 \\ \hline
\textbf{95-th}   & 7.95  & \textbf{4.38}  & 27.89 \\ \hline
\end{tabular}
\end{table}

\begin{table}[t]
\centering
\caption{Mean absolute errors of $\epsilon$-SVR and geometric methods for different sizes of the training set in cm}
\begin{tabular}{c|c|c|c|c}
\hline
\multicolumn{1}{c|}{\multirow{3}{*}{\textbf{\begin{tabular}[c]{@{}c@{}}Grid\\ size\end{tabular}}}} & \multirow{3}{*}{\textbf{\begin{tabular}[c]{@{}c@{}}Grid\\ resolution\end{tabular}}} & \multicolumn{3}{c}{\multirow{2}{*}{$\epsilon$-\textbf{SVR}}} \\\multicolumn{1}{c|}{}                                                            & & \multicolumn{3}{c}{}\\ \cline{3-5} \multicolumn{1}{c|}{}                                                                             &        & ULA     & DIS                  & URA          \\ \hline\hline
\multicolumn{1}{c|}{${\bf{6\times6}}$}
& 25.00  & 5.78    & \textbf{2.50}        & 16.06        \\ \hline
\multicolumn{1}{c|}{${\bf{11\times11}}$}
& 12.50  & 3.48    & \textbf{2.05}        & 13.95        \\ \hline
\multicolumn{1}{c|}{${\bf{26\times26}}$}                                                                
& 5.00   & 2.52    & \textbf{1.74}        & 9.97         \\ \hline
\multicolumn{1}{c|}{${\bf{51\times51}}$}                                                                
& 2.50   & 1.74    & \textbf{1.63}        & 9.16         \\ \hline\hline
\multicolumn{2}{c|}{\textbf{Triangulation-AoA}}                                                                                                                    & 8.06    & \textbf{7.57}   & 18.71        \\ \hline
\multicolumn{2}{c|}{\textbf{Trilateration-ToF}}                                                                                                                    & 60.55   & \textbf{49.58}  & 651.77          \\ \hline
\multicolumn{2}{c|}{\textbf{Trilateration-AMP}}                                                                                                                    & \textbf{229.01}  & 344.98          & 521.18          \\ \hline
\end{tabular}
\end{table}
\subsection{Impact of Antenna Topology}
We summarize the percentile errors of all-metric schemes in Fig. \ref{fig:SVR_metrics} in Table I. It can be observed that DIS achieves the highest accuracy, which has 1.88-cm median positioning errors and 4.38-cm 95-th percentile errors. Followed by URA with 9.93-cm median errors, ULA performs the second-highest median errors (2.67-cm) based on $\epsilon$-SVR. This phenomenon can be explained by the spatial-resolution difference of large-scale antenna arrays. The distributed array (DIS) is implemented surrounding the targeted area. Therefore, the angular information of all directions can obtain, whereas URA provides the smallest aperture and angular resolution among the three topologies.
\par
Table II further evaluates the impact of antenna topology by comparing the MAEs of the all-metric scheme with different sizes of fingerprints, namely $6\times6$, $11\times11$, $26\times26$, and $51\times51$. We also include the positioning errors of AoA-based triangulation, as well as ToF- and amplitude-based trilateration acting as the basic benchmarks. As shown in Table II, triangulation has much higher accuracy than ToF- or amplitude-based trilateration, which is consistent with the conclusion in Section V-A, namely, AoA outperforms the other two metrics for our bandwidth-limited massive MIMO system. As expected, the positioning accuracy of the proposed fingerprinting method exceeds triangulation. Table II shows that the URA has the most significant positioning errors among the three antenna topologies because of the worst spatial resolution among the three topologies. However, it can obtain a distinct accuracy improvement (from 16.06 cm to 9.16 cm) when increasing the size of the training set. Meanwhile, we observe that the training set's size slightly impacts SVR-based ULA and DIS topology. Even though trained with sparse grids (e.g., $6\times6$), ULA and DIS still achieve 5.78-cm and 2.5-cm MAEs, respectively.

\begin{figure}[t]
\centering
\setlength{\abovecaptionskip}{-0.02cm}
\includegraphics[width=0.32\textwidth]{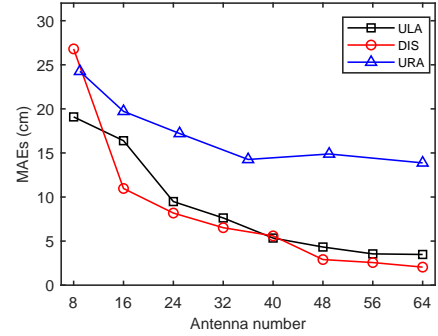}
\caption{Impact of number of antennas on positioning accuracy.}
\label{fig:PosErr_antSize}
\end{figure}
\subsection{Impact of Number of Antennas}
Fig. \ref{fig:PosErr_antSize} compares the positioning accuracy of the all-metric scheme in cases that different numbers of antenna have been utilized. For DIS and ULA, the number of the involved antenna elements varies from 8 to 64 with a step of 8. The selection of the number of antennas is based on the order of the antenna element is labeled in Fig. \ref{fig:MAMIMOSetup}. For URA, the antenna elements are selected as the square numbers (e.g., 9, 16, ..., 64) because we give the same weight on the azimuth and elevation. As shown in Fig. \ref{fig:PosErr_antSize}, the positioning errors generally decrease as the number of antennas increases but reach a plateau after a specific number of antennas. For example, the positioning accuracy of URA begins to saturate when the number of antennas reaches 36 (e.g., 6$\times$6) and fluctuates around 14 cm. Likewise, for ULA and DIS, the positioning errors only slightly decrease after the antenna's number hits 48.

\subsection{Impact of Signal-to-Noise Ratio}
\begin{figure}[t]
\centering
\setlength{\abovecaptionskip}{-0.01cm}
\subfigure[AoA estimation]{
\includegraphics[width=0.237\textwidth]{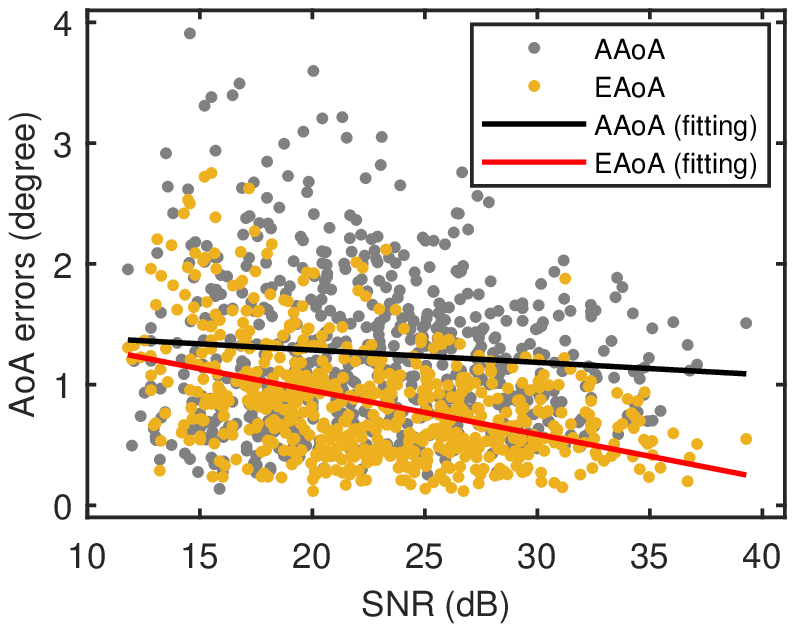}
}
\hspace{-4.9mm}
\subfigure[Positioning errors]{
\includegraphics[width=0.237\textwidth]{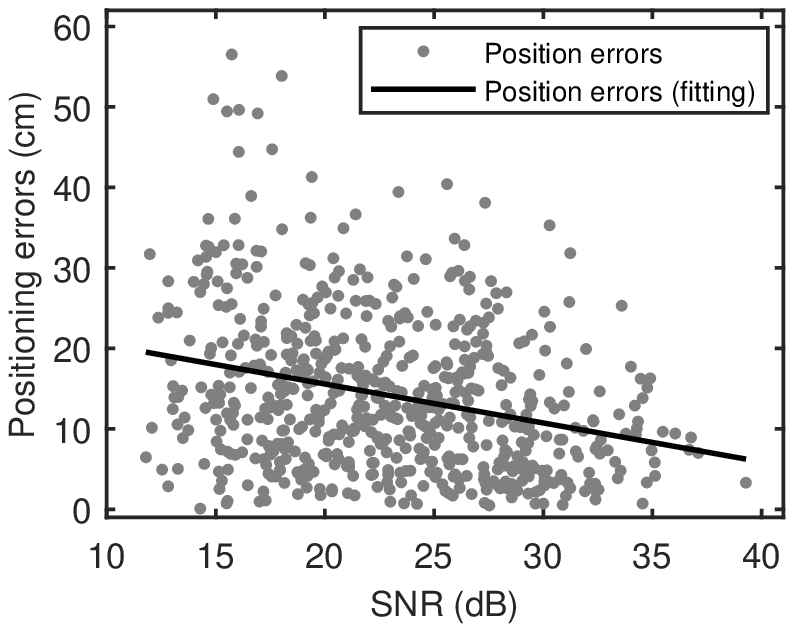}
}
\caption{Impact of SNR on angle-of-arrival estimation and positioning accuracy.}
\label{fig:URA_SNR}
\end{figure}

Generally, signal-to-noise ratio (SNR) can be calculated from RSSI and the noise measurement. However, the RSSI-based SNR may vary over packet reception and be easily affected by the frequency-selective fading due to multipath, which makes it difficult and inaccurate to evaluate system performance \cite{Halperin2010}. To assess the influence of SNR on positioning accuracy, the effective SNR for the OFDM system is introduced. According to \cite{Goldsmith2005}, the bit error rate (BER) $P_b$ is a function of the symbol SNR $\gamma$ for the narrow-band OFDM system. In our Massive MIMO system, QPSK modulation was adopted. So we have $P_b=Q(\sqrt{\gamma})$. The effective SNR was defined in \cite{Halperin2010} to give the same performance on the narrow-band channel despite frequency-selective fading and multiple streams (i.e., MIMO). Instead of simply average the SNR along the subcarriers, \cite{Halperin2010} proposed to average the subcarrier BERs and convert the mean BER to the effective SNR via the formula above. Since the transmitter antenna in our system is a single dipole, the BERs along the subcarriers can be calculated through \cite{Halperin2010},
\begin{equation}\label{eq:eff_SNR}
P_b=Q\left(\sqrt{\sum\mathop{}_{n_r}\mathbf{H}_{\rm{CSI}}\odot \mathbf{H}^{\ast}_{\rm{CSI}}}\right).
\end{equation}
The effective SNR $\gamma_{\rm{eff}}$ can be converted from the average BER by,
\begin{equation}\label{eq:eff_SNR}
\gamma_{\rm{eff}}=\left(Q^{-1}\left(\frac{1}{100}\sum\mathop{}_{n_k}P_b\right)\right)^2,
\end{equation}where 100 is the number of subcarriers. Note that for ULA and DIS in our experiment, the spatial scale of the antenna array is comparable to the size of the target area. So the SNR will vary a lot along the antenna array. In this subsection, we only consider the impact of effective SNR on URA. Fig. \ref{fig:URA_SNR}(a) presents the estimation errors of AoA versus the effective SNR. It is clear to observe that the errors of both azimuth and elevation decrease as the SNR increases, especially the elevation. So when converting to positioning errors based on $\epsilon$-SVR (with the 11$\times$11-grid fingerprints) in Section IV, as shown in Fig. \ref{fig:URA_SNR}(b), a similar tendency can be observed.

\subsection{Discussion and Future Works}
We have proposed an indoor fingerprinting method based on the MPCs of the direct links for a bandwidth-limited massive MIMO system, which localizes the UEs from the perspective of propagation and geometry. Compared with the CNN-based method in \cite{CNNMaMIMO2020} using the CSI matrix directly, the proposed fingerprinting method is generally adaptable to different scenarios and achieves a comparative positioning accuracy. Especially, the DIS topology achieves better accuracy with MAEs 1.63-2.5 cm compared with 8.23-cm MAEs in \cite{CNNMaMIMO2020}. However, the accuracy of URA (with MAEs 9.16-16.06 cm) is worse than the CNN-based method, which achieved 5.54-cm MAEs. The reason is two-fold. First, the CNN-based model has utilized much more extensive training sets (tens of thousand samples) than our method (144 to 10404 samples). Second, feeding CSI directly to CNN \cite{CNNMaMIMO2020} may learn more knowledge of multipath even though it is a black box. In this paper, we consider the direct link only and neglect the geometrical features from the reflection or scattering due to the strong LoS-scenario in our experiment. By considering those as well, multipath also helps to enhance spatial resolution. Based on the MPCs extracted from the FD-SAGE algorithm, we also can learn the geometry information of reflections and scattering links. The preliminary results in this paper pave the way for future multipath-assisted localization research.
\par
For the introduced massive MIMO systems, only nanosecond-level accuracy of ToF estimation was available due to the bandwidth limitation (20 MHz). But one nanosecond corresponds to 30-cm distance errors, which is insufficient for the centimeter-level (even millimeter-level) positioning. The intuitive solution is to have more bandwidth available, which is difficult (except for the millimeter wave) due to the spectrum scarcity in the sub-6 GHz band. In \cite{vasisht2015subNS}, a prototype Chronos was proposed to merge the measurement from several separated bands and stitch them together to give the illusion of a wideband radio. In this way, the authors achieved sub-nanosecond ToF accuracy using WiFi cards. Therefore, improving ToF accuracy based on the signal from the available hopping bands is also the potential direction to further enhance the positioning accuracy of the massive MIMO system. Moreover, recently, the phase-based distance metric was proposed for massive MIMO localization via channel sounder \cite{Li2019}. The phase has a high spatial resolution because a $2\pi$-phase shift corresponds to one-wavelength distance, which is quite promising for the fine-grained spatial resolution in case of limited bandwidth. However, the impact of the multipath effect on phase requires careful consideration for practical implementation.

\section{Conclusion}
This paper presents an indoor fingerprinting system based on a massive MIMO-OFDM testbed with a standard bandwidth. The raw CSI has been calibrated across the antenna and frequency domain using nonlinear regression. On top of that, the MPCs have been extracted based on the FD-SAGE algorithm. Exploiting the amplitude, AoA, and ToF from the direct links, we have implemented the indoor massive MIMO fingerprinting system based on $\epsilon$-SVR. The following results have been achieved.
\begin{enumerate}
\item We have investigated the positioning performance in the case of single-metric and hybrid-metric schemes. According to experimental validation, the AoA-based metric outperforms the other metrics in the bandwidth-limited massive MIMO system and dominates the positioning accuracy of any AoA-involved hybrid schemes.
\item We have achieved centimeter-level positioning errors generally with the relatively small training set. Especially, the DIS topology has achieved 2.50-1.63-cm MAEs regarding different sizes of the training set (namely, $6\times6$-grid to $51\times51$-grid for each UE).
\end{enumerate}
\par
For future work, embracing the MPCs from the reflection or scattering to enhance positioning performance will be investigated (e.g., the multipath-rich scenarios). Furthermore, other potential research directions are increasing the ToF accuracy based on the available bandwidth and exploiting the high-resolution distance metric (e.g., the phase).

\bibliographystyle{IEEEtran}
\bibliography{MaMIMOLoC}

\end{document}